\documentclass[12pt,a4paper]{article}
\usepackage{amsfonts,latexsym}
\usepackage{amsmath,amssymb}
\usepackage{graphicx,color}

\renewcommand{\thefootnote}{\fnsymbol{footnote}}

\begin{document}

\vspace{12mm}

\begin{center}
{{{\Large {\bf Primordial gravitational waves from conformal  gravity}}}}\\[10mm]

{Yun Soo Myung\footnote{e-mail address: ysmyung@inje.ac.kr} and Taeyoon Moon\footnote{e-mail address: tymoon@sogang.ac.kr}}\\[8mm]

{Institute of Basic Sciences and Department  of Computer Simulation, Inje University Gimhae 621-749, Korea\\[0pt]}

\end{center}
\vspace{2mm}

\begin{abstract}
We  investigate  the evolution of cosmological perturbations
generated during de Sitter inflation in the conformal gravity.
Primordial gravitational waves are composed of vector and  tensor
modes. We obtain  the constant vector and tensor  power spectra
which  seems to be correct   because the conformal gravity is
invariant under conformal transformation like the Maxwell kinetic
term.

\end{abstract}
\vspace{5mm}

{\footnotesize ~~~~PACS numbers: 98.80.Cq, 04.30.-w, 98.70.Vc }

{\footnotesize ~~~~Keywords: conformal gravity, power spectrum,
inflation}

\vspace{1.5cm}

\hspace{11.5cm}{Typeset Using \LaTeX}
\newpage
\renewcommand{\thefootnote}{\arabic{footnote}}
\setcounter{footnote}{0}


\section{Introduction}
The  detection of  primordial gravitational waves (GWs) via B-mode
polarization of Cosmic Microwave Background Radiation (CMBR) by
BICEP2~\cite{Ade:2014xna} has shown that the cosmic  inflation
occurred at a high scale of $10^{16}$ GeV is the most plausible
source of generating primordial GWs.  However, more data are
required to confirm  the above situation.  The primordial GWs can be
imprinted in the anisotropies and polarization spectrum of CMBR
 by making  the photon redshifts. The B-mode signal
observed by BICEP2 might  contain contributions from other sources
of vector modes and  cosmic strings,  in addition to tensor
modes~\cite{Moss:2014cra}.

The conformal gravity
$C^{\mu\nu\rho\sigma}C_{\mu\nu\rho\sigma}(=C^2)$ of being invariant
under the conformal transformation of $g_{\mu\nu}\to
\Omega^2(x)g_{\mu\nu}$ has its own interests in gravity and
cosmology. On the gravity side, it gives us  a promising combination
of $R_{\mu\nu}^2-R^2/3$ up to the Gauss-Bonnet term which kills
massive scalar GWs  when it couples to Einstein gravity (known to be
the Einstein-Weyl gravity)~\cite{Lu:2011zk}.
Stelle~\cite{Stelle:1976gc} has first introduced the quadratic
curvature gravity  of $a(R_{\mu\nu}^2-R^2/3)+b R^2$ to improve  the
perturbatively renormalizable property of Einstein gravity. In case
of $ab\not=0$, the renormalizability was achieved but the unitarity
was violated unless $a=0$, showing that the renormalizability and
unitarity exclude to each other.  Although the $a$-term of providing
massive GWs improves the ultraviolet divergence, it induces
simultaneously  ghost excitations which spoil the unitarity.  The
price one has to pay for making the theory renormalizable is the
loss of unitarity. This issue is not resolved completely until now.

However, the conformal gravity itself is renormalizable. Also, it
provides the AdS black hole solution~\cite{Riegert:1984zz} and its
thermodynamic properties and holography were discussed extensively
in the  literature~\cite{Klemm:1998kf,Lu:2012xu,Grumiller:2013mxa}.
The authors have investigated the AdS black hole thermodynamics and
stability in the Einstein-Weyl gravity and in the limit of  the
conformal gravity~\cite{Myung:2013uka}.

On the cosmology side of the conformal gravity, it provides surely a
massive vector propagation generated during de Sitter inflation in
addition to massive tensor ghosts when it couples to  Einstein
gravity~\cite{Clunan:2009er,Deruelle:2010kf,Deruelle:2012xv}.
Recently, the authors have shown that in the limit of $m^2 \to 0$
(keeping the conformal gravity only), the vector and tensor power
spectra disappear. It implies that their power spectra are not
gravitationally produced because the vector and tensor perturbations
are decoupled from the expanding de Sitter background. This occurs
due to conformal invariance as a transversely  massive vector has
been shown in the $m^2\to 0$ limit of the massive Maxwell theory
($-F^2/4+m^2A^2/2$)~\cite{Dimopoulos:2006ms}. We note here that
$F^2$ is conformally invariant  like $C^2$ under the transformation
of  $g_{\mu\nu}\to \Omega^2(x)g_{\mu\nu}$~\cite{Myung:2014aia}.
 The
conformal gravity implication to cosmological perturbation was first
studied in~\cite{Mannheim:2011is} which might indicate  that there
exists a difference between conformal and Einstein gravities in
their perturbed equations around de Sitter background. Even though
he has obtained a ``degenerate fourth-order equation" for the metric
perturbation tensor $h_{\mu\nu}$ from the conformal gravity,  any
relevant quantity was not found because he did not split
$h_{\mu\nu}$ according to the SO(3) decomposition for cosmological
perturbations. As far as we know, there is no definite computation
of an observable like the power spectrum in the conformal gravity.

In this Letter, we will study the conformal  gravity  as a
higher-order gravity theory to compute the vector and tensor power
spectra generated from de Sitter inflation. Considering the
conformal invariant of the conformal gravity seriously, we expect to
obtain the constant power spectra for vector and tensor
perturbations.

\section{Conformal  gravity }

Let us first consider the conformal  gravity whose action is given
by
\begin{equation} \label{ECSW}
S_{\rm CG}=\frac{1}{4\kappa m^2}\int d^4x
\sqrt{-g}\Big[C^{\mu\nu\rho\sigma}C_{\mu\nu\rho\sigma}\Big],
\end{equation}
where the Weyl-squared term is given by
\begin{eqnarray}
C^{\mu\nu\rho\sigma}C_{\mu\nu\rho\sigma}&=&2\Big(R^{\mu\nu}R_{\mu\nu}-\frac{1}{3}R^2\Big)+
(R^{\mu\nu\rho\sigma}R_{\mu\nu\rho\sigma}-4R^{\mu\nu}R_{\mu\nu}+R^2)
\end{eqnarray}
with the Weyl tensor
\begin{equation}
C_{\mu\nu\rho\sigma}=R_{\mu\nu\rho\sigma}-\frac{1}{2}\Big(
g_{\mu\rho}R_{\nu\sigma}-g_{\mu\sigma}R_{\nu\rho}-g_{\nu\rho}R_{\mu\sigma}+g_{\nu\sigma}R_{\mu\rho}\Big)+\frac{1}{6}R(g_{\mu\rho}g_{\nu\sigma}-g_{\mu\sigma}g_{\nu\rho}).
\end{equation}
 Here we have
$\kappa=8\pi G=1/M^2_{\rm P}$, $M_{\rm P}$ being the reduced Planck
mass and a mass-squared $m^2$ is introduced to make the action
dimensionless. Greek indices run from 0 to 3 with conventions
$(-+++)$, while Latin indices run from 1 to 3. Further, we note that
the Weyl-squared term is invariant under the conformal
transformation of $g_{\mu\nu} \to \Omega^2(x) g_{\mu\nu}$.

Its  equation takes the form
\begin{equation} \label{ein-eq}
2 \nabla^\rho\nabla^\sigma
C_{\mu\rho\nu\sigma}+G^{\rho\sigma}C_{\mu\rho\nu\sigma}=0
\end{equation} with  the Einstein tensor $G_{\mu\nu}$.
The solution is de Sitter space whose  curvature quantities are
given by
\begin{equation}
\bar{R}_{\mu\nu\rho\sigma}=H^2(\bar{g}_{\mu\rho}\bar{g}_{\nu\sigma}-\bar{g}_{\mu\sigma}\bar{g}_{\nu\rho}),~~\bar{R}_{\mu\nu}=3H^2\bar{g}_{\mu\nu},~~\bar{R}=12H^2
\end{equation}
with $H$=constant. We choose  de Sitter  background explicitly  by
choosing a conformal time $\eta$
\begin{eqnarray} \label{frw}
ds^2_{\rm dS}=\bar{g}_{\mu\nu}dx^\mu
dx^\nu=a(\eta)^2\Big[-d\eta^2+\delta_{ij}dx^idx^j\Big],
\end{eqnarray}
where the conformal  scale factor is
\begin{eqnarray}
a(\eta)=-\frac{1}{H\eta}\to a(t)=e^{Ht}.
\end{eqnarray}
Here the latter denotes  the scale factor with respect to cosmic
time $t$.

We   choose  the Newtonian gauge of $B=E=0 $ and $\bar{E}_i=0$ which
leads to $10-4=6$ degrees of freedom (DOF). In this case, the
cosmologically perturbed metric can be simplified to be
\begin{eqnarray} \label{so3-met}
ds^2=a(\eta)^2\Big[-(1+2\Psi)d\eta^2+2\Psi_i d\eta
dx^{i}+\Big\{(1+2\Phi)\delta_{ij}+h_{ij}\Big\}dx^idx^j\Big]
\end{eqnarray}
with the transverse vector $\partial_i\Psi^i=0$ and
transverse-traceless tensor $\partial_ih^{ij}=h=0$. We emphasize
that choosing the SO(3)-perturbed metric (\ref{so3-met}) contrasts
sharply with the covariant approach to the  cosmological conformal
gravity~\cite{Mannheim:2011is}.

In order to get  the cosmological perturbed equations,  one  is
first to obtain the bilinear action and then, varying it to yield
the perturbed equations.  We expand the conformal gravity action
(\ref{ECSW}) up to quadratic order in the perturbations of
$\Psi,\Phi,\Psi_i,$ and $h_{ij}$ around  the de Sitter
background~\cite{Deruelle:2010kf}.  Then,  the bilinear  actions for
scalar, vector and tensor perturbations can be found as
\begin{eqnarray}
&&\hspace*{-2.3em}S_{\rm CG}^{({\rm S})}=\frac{1}{3\kappa m^2}\int
d^4x\Big[\nabla^2 (\Psi-\Phi)\Big]^2,
\label{scalar}\\
&&\nonumber\\
 &&\hspace*{-2.3em}S_{\rm CG}^{({\rm V})}=\frac{1}{4\kappa m^2}\int
d^4x\Big(\partial_i\Psi'_{j}\partial^{i}\Psi'^{j}
-\nabla^2\Psi_i\nabla^2\Psi^i\Big),\label{vpeq}\\
&&\nonumber\\
&&\hspace*{-2.3em} S_{\rm CG}^{({\rm T})}=\frac{1}{8\kappa m^2}\int
d^4x\Big(h''_{ij}h''^{ij} -2\partial_kh'_{ij}\partial^{k}h'^{ij}
+\nabla^2h_{ij}\nabla^2h^{ij}\Big)\label{hpeq}.
\end{eqnarray}
Varying the actions (\ref{vpeq}) and (\ref{hpeq}) with respect to
$\Psi^{i}$ and $h^{ij}$  leads to the equations of motion for vector
and tensor perturbations
\begin{eqnarray}
&&\Box\Psi_i=0,\label{veq}\\
 &&\Box^2h_{ij}=0,\label{heq}
\end{eqnarray}
where $\Box=d^2/d\eta^2-\nabla^2$ with $\nabla^2$ the Laplacian
operator.
 It is worth noting  that  Eqs.(\ref{veq}) and (\ref{heq})  are independent of
the expanding de Sitter background in the conformal  gravity.

Finally, we would like to mention two scalars $\Phi$ and $\Phi$. Two
scalar equations
 are given by $\nabla^2\Psi=\nabla^2\Phi=0$,  which implies
that they are obviously
 non-propagating modes  in the de Sitter background. This means that the cosmological conformal gravity describes 4 DOF of vector and tensor modes. Hereafter,
 thus, we will not consider the scalar sector.

\section{Primordial power spectra}
The power spectrum is  given by the two-point correlation function
which could be  computed when one chooses   the vacuum state
$|0\rangle$. It is defined by
\begin{equation}
\langle0|{\cal F}(\eta,\bold{x}){\cal
F}(\eta,\bold{x}')|0\rangle=\int d^3\bold{k} \frac{{\cal P}_{\cal
F}}{4\pi k^3}e^{-i \bold{k}\cdot (\bold{x}-\bold{x}')},
\end{equation}
where ${\cal F}$ denotes vector and tensor and $k=|\bold{k}|$ is the
wave number. In general, fluctuations are created on all length
scales with wave number $k$. Cosmologically relevant fluctuations
start their lives inside the Hubble radius which defines the
subhorizon: $k~\gg aH~(z=-k\eta\gg 1)$.  On the other hand, the
comoving Hubble radius $(aH)^{-1}$ shrinks during inflation while
the comoving wavenumber $k$ is constant. Therefore, eventually all
fluctuations exit the comoving  Hubble radius which defines the
superhorizon: $k~\ll aH~(z=-k\eta\ll 1)$.

One may compute the two-point function by taking the Bunch-Davies
vacuum $|0\rangle$.
 In the de Sitter inflation, we choose the subhorizon limit
of  $z\to \infty$  to define the Bunch-Davies vacuum, while we
choose the superhorizon limit of $z\to 0$ to get a definite form of
the power spectrum which stays alive after decaying.  For example,
fluctuations of scalar and tensor originate on subhorizon scales and
they propagate for a long time on superhorizon scales. This can be
checked by computing their power spectra which are scale-invariant.
Accordingly, it would be interesting  to check what happens when one
computes the power spectra for vector and tensor perturbations
generated from de Sitter inflation in the frame work of conformal
gravity.

\subsection{Vector power spectrum}

Let us  consider Eq.(\ref{veq}) for vector perturbation and then,
expand  $\Psi_i$ in plane waves with the linearly polarized states
\begin{eqnarray}\label{psim}
\Psi_i(\eta,{\bf x})=\frac{1}{(2\pi)^{\frac{3}{2}}}\int d^3{\bf
k}\sum_{s=1,2}p_i^{s}({\bf k})\Psi_{\bf k}^{s}(\eta)e^{i{\bf
k}\cdot{\bf x}},
\end{eqnarray}
where  $p^{1/2}_{i}$ are linear polarization   vectors with
 $p^{1/2}_i
p^{1/2, i}=1$.  Also, $\Psi_{\bf k}^{s}$ denote linearly polarized
vector modes. Plugging (\ref{psim}) into the equation (\ref{veq}),
one finds the equation
\begin{eqnarray}\label{v0eq}
\Bigg[\frac{d^2}{d\eta^2}+k^2\Bigg]\Psi_{\bf k}^s(\eta)=0.
\end{eqnarray}
Introducing  $z=-k\eta$, Eq.(\ref{v0eq}) takes a simple form
\begin{equation}\label{v1eq}
\Big[\frac{d^2}{dz^2}+1\Big]\Psi_{\bf k}^{s}(z)=0\end{equation}
whose positive frequency solution is given by \begin{equation}
\Psi_{\bf k}^{s}(z)\sim e^{iz}\end{equation} up to the
normalization.

 We are willing to  calculate
vector power spectrum. For this purpose, we define a commutation
relation for the vector. In the bilinear action (\ref{vpeq}), the
conjugate momentum for the field $\Psi_j$ is found to be
\begin{eqnarray}\label{vconj}
\pi_{\Psi}^{j}=-\frac{1}{2\kappa m^2}\nabla^2\Psi'^{j},
\end{eqnarray}
where one observes an unusual factor $\nabla^2$ which reflects that
the vector $\Psi_i$ is not a canonically defined vector, but it is
from the cosmological conformal gravity.  The canonical quantization
is implemented by imposing the commutation relation
\begin{eqnarray}\label{vcomm}
[\hat{\Psi}_{j}(\eta,{\bf x}),\hat{\pi}_{\Psi}^{j}(\eta,{\bf
x}^{\prime})]=2i\delta({\bf x}-{\bf x}^{\prime})
\end{eqnarray}
with $\hbar=1$.

Now, the operator $\hat{\Psi}_{j}$ can be expanded in Fourier modes
as
\begin{eqnarray}\label{vex}
\hat{\Psi}_{j}(\eta,{\bf x})=\frac{1}{(2\pi)^{\frac{3}{2}}}\int
d^3{\bf k}\sum_{s=1,2}\Big(p_{j}^{s}({\bf k})\hat{a}_{\bf
k}^{s}\Psi_{\bf k}^{s}(\eta)e^{i{\bf k}\cdot{\bf x}}+{\rm h.c.}\Big)
\end{eqnarray}
and the operator $\hat{\pi}_{\Psi}^{j}=\frac{k^2}{2\kappa
m^2}\hat{\Psi}'^{j}$ can be easily obtained from (\ref{vex}).
Plugging (\ref{vex}) and $\hat{\pi}_{\Psi}^{j}$ into (\ref{vcomm}),
we find the commutation relation and Wronskian condition as
\begin{eqnarray}
&&\hspace*{-2em}[\hat{a}_{\bf k}^{s},\hat{a}_{\bf k^{\prime}}^{
s^{\prime}\dag}]=\delta^{ss^{\prime}}\delta^3({\bf k}-{\bf
k}^{\prime}),\label{comm0}\\
&&\hspace*{-2em}\Psi_{\bf k}^{s}\Big(\frac{k^2}{2\kappa
m^2}\Big)(\Psi_{\bf k}^{*s})^{\prime}-{\rm c.c.}=i \to \Psi_{\bf
k}^{s}\frac{d\Psi_{\bf k}^{*s}}{dz}-{\rm c.c.}=-\frac{2i\kappa
m^2}{k^3}. \label{vwcon}
\end{eqnarray}
  We
choose the  positive frequency mode for a Bunch-Davies vacuum
$|0\rangle$ normalized by the Wronskian condition
\begin{eqnarray} \label{vecsol}
\Psi_{\bf k}^{s}(z) =\sqrt{\frac{\kappa m^2}{k^3}} e^{iz}
\end{eqnarray}
as the solution to (\ref{v1eq}).
 On the other hand, the
vector power spectrum is defined by
\begin{eqnarray}\label{powerv}
\langle0|\hat{\Psi}_{j}(\eta,{\bf x})\hat{\Psi}^{j}(\eta,{\bf
x}')|0\rangle=\int d^3{\bf k}\frac{{\cal P}_{\Psi}}{4\pi
k^3}e^{i{\bf k}\cdot({\bf x}-{\bf x^{\prime}})},
\end{eqnarray}
where we take  the Bunch-Davies vacuum state $|0\rangle$  by
imposing $\hat{a}_{\bf k}^{s}|0\rangle=0$. The vector  power
spectrum ${\cal P}_{\Psi}$ takes the form  \begin{equation}
\label{vecpt}{\cal
P}_{\Psi}\equiv\sum_{s=1,2}\frac{k^3}{2\pi^2}\Big|\Psi_{\bf
k}^{s}\Big|^2.\end{equation}
 Plugging (\ref{vecsol}) into
(\ref{vecpt}), we find a constant power  spectrum for a vector
perturbation
\begin{eqnarray} \label{vec-pow}
{\cal P}_{\Psi}=\frac{m^2}{\pi^2 M^2_{\rm P}}.
\end{eqnarray}

\subsection{Tensor power spectrum}

We take   Eq.(\ref{heq}) to compute  tensor power spectrum. In this
case, the metric tensor $h_{ij}$ can be expanded in Fourier modes
\begin{eqnarray}\label{hijm}
h_{ij}(\eta,{\bf x})=\frac{1}{(2\pi)^{\frac{3}{2}}}\int d^3{\bf
k}\sum_{s={\rm +,\times}}p_{ij}^{s}({\bf k})h_{\bf
k}^{s}(\eta)e^{i{\bf k}\cdot{\bf x}},
\end{eqnarray}
where  $p^{s}_{ij}$ linear polarization tensors with $p^{s}_{ij}
p^{s,ij}=1$. Also, $h_{\bf k}^{s}(\eta)$ represent linearly
polarized tensor modes. Plugging (\ref{hijm}) into (\ref{heq}) leads
to the fourth-order differential equation
\begin{eqnarray}
&&(h_{\bf k}^{s})^{''''}+2k^2(h_{\bf k}^{s})^{''}+k^4h_{\bf k}^{s}
=0,\label{heq2}
\end{eqnarray}
which is further rewritten as a factorized form
\begin{eqnarray}
&&\Bigg[\frac{d^2}{d\eta^2}+k^2\Bigg]^2h_{\bf k}^{s}(\eta)=0.
\label{hee}
\end{eqnarray}
Introducing  $z=-k\eta$, Eq.(\ref{hee}) takes the form of a
degenerate fourth-order equation
\begin{equation}\label{hc0}
\Big[\frac{d^2}{dz^2}+1\Big]^2h_{\bf k}^{s}(z)=0.\end{equation} This
is  the same equation for a degenerate Pais-Uhlenbeck (PU)
oscillator and its solution is given by
\begin{equation} \label{desol}
h_{\bf k}^{s}(z)=\frac{N}{2k^2}\Big[(a_2^s+a_1^s z)e^{iz}+c.c.\Big]
\end{equation}
 with $N$ the normalization constant. After quantization, $a^s_2$
 and $a^s_1$ are promoted to operators $\hat{a}^s_2({\bf k})$ and $\hat{a}^s_1({\bf
 k})$.   The presence of $z$
in $(\cdots)$ reflects clearly that $ h_{\bf k}^{s}(z)$ is a
solution to the degenerate
 equation (\ref{hc0}).
However, it is difficult to quantize $h_{ij}$ in the subhorizon
region directly  because it satisfies the degenerate fourth-order
equation (\ref{heq}). In order to quantize $h_{ij}$, we have to
consider (\ref{heq}) as a final equation obtained by making use of
an auxiliary tensor $\beta_{ij}$.

For this purpose,  one rewrites the fourth-order action (\ref{hpeq})
as a second-order action
\begin{equation}
S_{\rm AC}^{({\rm T})}=-\frac{1}{4 \kappa m^2}\int
d^4x\Big(\eta^{\mu\nu}\partial_\mu \beta_{ij}\partial_\nu h^{ij}
+\frac{1}{2} \beta_{ij}\beta^{ij}\Big).\label{ahpeq}
\end{equation}
Their equations are given by
\begin{equation}
\Box h_{ij}=\beta_{ij},~~\Box \beta_{ij}=0,
\end{equation}
which are combined to give the fourth-order tensor equation
(\ref{heq}). Explicitly, acting $\Box$ on the first equation leads
to (\ref{heq}) when one uses the second one. Actually, this is an
extension of the singleton  action describing a dipole ghost pair as
the fourth-order scalar
theory~\cite{Myung:1999nd,Rivelles:2003jd,Kim:2013waf,Kim:2013mfa}.
This is related to not a non-degenerate PU oscillator and its
quantization, but a degenerate PU and
quantization~\cite{Mannheim:2004qz,Smilga:2008pr}. The canonical
conjugate momenta are given by
\begin{equation}
\pi_h^{ij}=\frac{1}{4\kappa
m^2}\beta'^{ij},~~\pi_\beta^{ij}=\frac{1}{4\kappa m^2}h'^{ij}.
\end{equation}
After expanding $\hat{h}_{ij}$ and $\hat{\beta}_{ij}$ in their
Fourier modes, their amplitudes at each mode are given by
\begin{eqnarray}
\label{sol1}\hat{\beta}^s_{\bf k}(z)&=&iN\Big(\hat{a}^s_1({\bf k})e^{iz}-\hat{a}^{s\dagger}_1({\bf k})e^{-iz}\Big),\\
 \label{sol2}\hat{h}^s_{\bf k}(z)&=&\frac{N}{2k^2}\Big[\left(\hat{a}^s_2({\bf k})+\hat{a}^s_1({\bf
 k})z\right)e^{iz}+{\rm h.c.}\Big].
 \end{eqnarray}
Now, the canonical quantization is accomplished by imposing
equal-time commutation relations:
\begin{eqnarray}\label{comm}
[\hat{h}_{ij}(\eta,{\bf x}),\hat{\pi}_{h}^{ij}(\eta,{\bf
x}^{\prime})]=2i\delta^3({\bf x}-{\bf
x}^{\prime}),~~[\hat{\beta}_{ij}(\eta,{\bf
x}),\hat{\pi}_{\beta}^{ij}(\eta,{\bf x}^{\prime})]=2i\delta^3({\bf
x}-{\bf x}^{\prime}),
\end{eqnarray}
where the factor 2 is coming  from the fact that $h_{ij}$ and
$\beta_{ij}$ represent 2 DOF, respectively. Taking (\ref{sol1}) and
(\ref{sol2}) into account, the two operators $\hat{\beta}_{ij}$ and
$\hat{h}_{ij}$ are given by
\begin{eqnarray}\label{hex1}
\hat{\beta}_{ij}(z,{\bf x})&=&\frac{ N}{(2\pi)^{\frac{3}{2}}}\int
d^3{\bf k}\Bigg[\sum_{s=+,\times}\Big(ip_{ij}^{s}({\bf
k})\hat{a}_1^s({\bf k})e^{iz}e^{i{\bf k}\cdot{\bf
x}}\Big)+{\rm h.c.}\Bigg], \\
\label{hex2} \hat{h}_{ij}(z,{\bf
x})&=&\frac{1}{(2\pi)^{\frac{3}{2}}}\int d^3{\bf
k}\frac{N}{2k^2}\Bigg[\sum_{s=+,\times}\Big\{p_{ij}^{s}({\bf
k})\Big(\hat{a}_2^s({\bf k})+\hat{a}_1^s({\bf k})z
\Big)e^{iz}e^{i{\bf k}\cdot{\bf x}}\Big\}+{\rm h.c.}\Bigg].
\end{eqnarray}
Plugging (\ref{hex1}) and (\ref{hex2}) into (\ref{comm}) determines
the normalization constant $N=\sqrt{2\kappa m^2}$ and commutation
relations between $\hat{a}_i^s({\bf k})$ and
$\hat{a}^{s\dagger}_j({\bf k}')$ as
 \begin{equation} \label{scft}
 [\hat{a}_i^s({\bf k}), \hat{a}^{s^{\prime}\dagger}_j({\bf k}')]= 2k \delta^{ss'}
 \left(
  \begin{array}{cc}
   0 & -i  \\
    i & 1 \\
  \end{array}
 \right)\delta^3({\bf k}-{\bf k}').
 \end{equation}
Here the commutation  relation of $ [\hat{a}_2^s({\bf k}),
\hat{a}^{s^{\prime}\dagger}_2({\bf k}')]$ is determined by the
condition of  \begin{equation} [\hat{h}_{ij}(\eta,{\bf
x}),\hat{\pi}_{\beta}^{ij}(\eta,{\bf x}^{\prime})]=0.
\end{equation}
We are ready to compute  the power spectrum of the
gravitational waves defined by
\begin{eqnarray}\label{power}
\langle0|\hat{h}_{ij}(\eta,{\bf x})\hat{h}^{ij}(\eta,{\bf
x^{\prime}})|0\rangle=\int d^3k\frac{{\cal P}_{\rm h}}{4\pi
k^3}e^{i{\bf k}\cdot({\bf x}-{\bf x^{\prime}})}.
\end{eqnarray}
Here we choose the Bunch-Davies vacuum $|0\rangle$ by imposing
$\hat{a}_i^s({\bf k})|0\rangle=0$.  The tensor power spectrum ${\cal
P}_{\rm h}$ in (\ref{power}) denotes ${\cal P}_{\rm
h}\equiv\sum_{s={+,\times}}{\cal P}^s_{\rm h}$ where $ {\cal
P}^s_{\rm h}$ is given as
\begin{eqnarray}
{\cal P}^s_{\rm h}=\frac{k^3}{2\pi^2}|h_{\bf
k}^{s}\Big|^2=\frac{m^2}{2\pi^2M^2_{\rm P}}.
\end{eqnarray}
Finally, we obtain the tensor power spectrum
\begin{equation}
{\cal P}_{\rm h}=\frac{m^2}{\pi^2M^2_{\rm P}}
\end{equation}
which corresponds to a constant power spectrum. This is the same
form as for the vector power spectrum (\ref{vec-pow}).

On the other hand, the power spectrum of auxiliary tensor
$\beta_{ij}$ is  defined  by
\begin{equation}
\langle0|\hat{\beta}_{ij}(\eta,{\bf x})\hat{\beta}^{ij}(\eta,{\bf
x^{\prime}})|0\rangle=\int d^3k\frac{{\cal P}_{\rm \beta}}{4\pi
k^3}e^{i{\bf k}\cdot({\bf x}-{\bf x^{\prime}})}. \end{equation} Here
we obtain the zero power spectrum as
\begin{equation}
{\cal P}_{\rm \beta}=0
\end{equation}
when one used  the commutation relation $[\hat{a}_1^s({\bf k}),
\hat{a}^{s^{\prime}\dagger}_1({\bf k}')]=0$. This is clear because
$\beta_{ij}$ is an auxiliary tensor to lower the fourth-order action
to the second-order action. However, it is not  understand why
$\hat{h}_{ij}$ could be expanded  by $\hat{a}_2^s({\bf k})$ and
$\hat{a}_1^s({\bf k})$ without introducing  $\beta_{ij}$, because
$\beta_{ij}$ was expanded by $\hat{a}_1^s({\bf k})$ solely.

\section{Discussions}

We have found the constant vector and tensor power spectra generated
during de Sitter inflation from conformal gravity. These constant
power spectra could be understood because the conformal gravity is
invariant under conformal (Weyl) transformation. This means that
their power spectra are constant with respect to $z=-k\eta$ since
vector and tensor perturbations are decoupled from the expanding de
Sitter inflation. In other words, this is so because the bilinear
actions (\ref{vpeq}) and (\ref{hpeq}) are independent of the
conformal scale factor $a(\eta)$ as a result of conformal
invariance. On the contrary of Ref.\cite{Mannheim:2011is},  it is
less interesting for the conformal gravity to further investigate
its cosmological implications.

 Hence, our
analysis implies that the Einstein-Weyl gravity is more promising
than the conformal gravity to obtain the physical  tensor power
spectrum because the Einstein-Hilbert term provides the coupling of
scale factor $a$ like as
$a^2(h'_{ij}h'^{ij}-\partial_lh_{ij}\partial^lh^{ij})$. Also, the
singleton Lagrangian of ${\cal
L}_s=-\sqrt{-\bar{g}}(\frac{1}{2}\bar{g}^{\mu\nu}\partial_\mu\phi_1\partial_\nu\phi_2+\frac{1}{2}\phi_1^2)$
is quite interesting because it provides two scalar equations $(\Box
+2aH\partial_\eta)\phi_2=\phi_1$ and $(\Box +2aH\partial_\eta)
\phi_1=0$ which are combined to yield the degenerate fourth-order
equation $(\Box +2aH\partial_\eta)^2 \phi_2=0$. Here we observe the
presence of the scale factor $a$  in the perturbed equation of the
singleton.

 Consequently, the
conformal invariance of the Lagrangian  like  $\sqrt{-g}C^2$ or
$\sqrt{-g}F^2$ has no responsibility for generating the observed
fluctuations during inflation.

\vspace{0.25cm}

 {\bf Acknowledgement}

\vspace{0.25cm}
 This work was supported by the National
Research Foundation of Korea (NRF) grant funded by the Korea
government (MEST) (No.2012-R1A1A2A10040499).

\end{document}